\documentclass[aps,pra,reprint,superscriptaddress]{revtex4-1}

\usepackage{lineno}

\usepackage{hyperref}
\usepackage{graphicx}  
\usepackage{dcolumn}   
\usepackage{bm}        
\usepackage{amssymb}   
\usepackage{epstopdf}
\usepackage{amsmath}
\usepackage{xspace}
\usepackage{verbatim}
\usepackage{color}
\usepackage{xfrac}

\usepackage{soul}

\newcommand{\ket}[1]{\left| #1 \right\rangle}

\hypersetup{
    colorlinks=true,linkcolor=blue,citecolor=blue,
    filecolor=blue,urlcolor=blue,breaklinks=true
}

\begin{document}
\title{Particle states are equidistant to wave and fully-entangled states in an interferometer }

\author{Miguel Orszag}
\email{morszag@fis.puc.cl}
\affiliation{Instituto de F\'{i}sica, Pontificia Universidad Cat\'{o}lica de Chile,
Casilla 306, Santiago, Chile}
\affiliation{Centro de \'{O}ptica e Informaci\'{o}n Cu\'{a}ntica, Camino la Pir\'{a}mide 5750, Huechuraba, Santiago, Chile}

\author{Sergio Carrasco}
\affiliation{Instituto de F\'{i}sica, Pontificia Universidad Cat\'{o}lica de Chile,
Casilla 306, Santiago, Chile}

\date{\today}

\begin{abstract}
In this article we show that, in a two-arm interferometer, pure quantum states of perfect path distinguishability (particles) are geometrically equidistant from all states with constant path distinguishability $D$. This property is not shared by other states, such as perfect fringe-visibility (waves) or maximally entangled quantum states (entanglon).  Indeed, the Bures distance between a particle and any other state depends only the distinguishability of the latter. On the contrary, the Bures distance between a wave or an entanglon, and any other single photon state depends on other set of parameters.  
\end{abstract}

\maketitle

\section{Introduction}

Bohr introduced the \textit{principle of complementarity} \cite{Bohr,ScullyEnglertWalther} shortly after the famous paper by Einstein, Podoslky and Rosen (EPR) \cite{EPR}. This principle states that results of two mutually incompatible tests can not be \textit{jointly} taken into account, in order to establish conclusions regarding the ``elements of physical reality'' of a given system. 

As an example, consider the entangled state given by EPR. In this case, the result of a measurement of the position of the first particle, allows to predict with certainty the position of a second, distant particle. Thus, the position of the second particle is an element of physical reality (first conclusion). On the other hand, if the momentum of the first particle is measured, then the momentum of the second particle is also an element of physical reality (second conclusion). By respecting locality, i.e. a system can not be affected a by another distant system, EPR concluded that Quantum Mechanics is either inconsistent (it leads to contradictions, since momentum and position are not simultaneously determined) or incomplete. 

Bohr, however, by appealing to the principle of complementary, argued that both conclusions should be taken separately since they are obtained from two different \textit{incompatible experimental configurations}. 

Therefore, quantum systems have real properties that are mutually exclusive, the best example being what is known as wave-particle duality. In a two-way interferometer, the wavelike property is associated to the interference fringes, whereas the particlelike behaviour is related to the which-way information (WWI). The first is quantified by the interferometric visibility $V$ and the latter by different kind of measures.

Greenberg and Yasin \cite{GY} employed the predictability $P$ to quantify the WWI, a real number ranging from 0 (no WWI) to 1 (full WWI). In a two-path interferometer, the predictability of a single photon in a pure state can be defined as
\begin{eqnarray}
P=||\langle 0| \psi \rangle |^2 - |\langle 1| \psi \rangle |^2|,
\end{eqnarray}
where $\ket{0}$ and $\ket{1}$ are single photon states, indicating one photon in mode-0 and one photon in mode-1, each mode representing a path of the interferometer. Therefore, $\langle 0 | 1 \rangle =0$.  They demonstrated the inequality 
\begin{eqnarray}\label{GY}
P^2+V^2\leq1,
\end{eqnarray}  
which quantifies the the wave-particle duality, and shows that different levels of wave and particle properties may be simultaneously measured. For interferometers with more than two paths,  Jaeger et al. \cite{Jaeger} proposed other definitions of path information and visibility to formulate interferometric complementarities. 

Based on the works of Zurek and Wooters \cite{ZW}, Englert \cite{Englert3,ENG,Englert} added a which way detector (WWD) on each arm of the interferometer and quantified the WWI by defining the path distinguishability  $D$  as 
\begin{eqnarray}\label{D}
D= \frac{1}{2} Tr \{  | \rho_{D}^{0} - \rho_{D}^{1} | \}
\end{eqnarray}
while the fringe visibility was modified by the presence of the WWD according to the expression
\begin{eqnarray}\label{Vis}
V=|  Tr\{    \hat{U}_0^{\dagger} \rho_D^{(i)}  \hat{U}_{1} \} |. 
\end{eqnarray}
In these expressions, $\hat{U}_{0}$ and $\hat{U}_{0}$ are unitary operators that locally entangle the photon to the WWD in the arms 0 and 1, respectively,  $\rho_{D}^{(i)}$ is the initial state of the detector, and $\rho_{D}^{0,1}=\hat{U}_{0,1}^{\dagger}\rho_{D}^{i}\hat{U}_{0,1}$. Englert showed that these quantities obey a similar inequality to (\ref{GY}), namely
\begin{eqnarray}\label{EnglertIN}
V^2+D^2\leq1.
 \end{eqnarray} 
 
Notice that inequality  (\ref{GY}) establishes a relation between single-particle properties. On the contrary, in (\ref{EnglertIN}) the visibility and distinguishability are defined in terms of a second system, the which way detector. 
 
In \cite{Bergou1} entanglement was taken into account, leading to relations between single-particle properties (visibility and distinguishability) and the concurrence, which is a bipartite property with no classical analog (a measure of etanglement). 
In \cite{Bergou1,Bergou2}  two coherence measures (an $l_1$ and an entropic measure) were introduced to establish relations between which-path information and coherence, for an interferometer with multiple paths.

On the other hand, for both, classical and quantum light, the polarization coherence theorem (PCT) \cite{EberlyPCT,EberlyPCT2,EberlyPCT3,DeZela1,DeZela2} joins polarization $(P)$ to wave-ray duality through a tight equality $V^2+D^2=P^2$.  Recently, for single-photons, Qian \textit{et al. } \cite{Qian} employed the concurrence $C$, a measure of quantum entanglement between \textit{two degrees of freedom of a single photon}, to show that 
\begin{eqnarray}\label{Qian}
V^2+D^2+C^2=1.
\end{eqnarray}
This result indicates that both, particle and wave properties of a single photon, may be simultaneously turned off. Within the possible quantons defined by this relation, we define waves ($D=1$), particles ($V=1$) and ``entanglons'' ($C=1$).
The structure of this article is as follows. In section \ref{sec:sec2} we characterise a family of single photon states in terms of its $D$, $V$ and $C$ values. In section \ref{sec:sec3}, the Bures distance between two arbitraty families of states is calculated. Then, we show that a particle state ($D=1$) has a unique symmetry, as far as distance is concerned, with respect to the wave or entanglon.  This constitutes the main result of or article, which is further commented in section \ref{sec:sec4}.  

\section{General 2-qubit states}\label{sec:sec2} 

Let us consider a two-arm interferometer, whose arms are labeled by 0 and 1, and define a 2-qubit state $\ket{\psi} $ of a first quantum system, or \textit{quanton}, as follows 
\begin{eqnarray}
\ket{\psi}&=&\sqrt{\frac{1+ D}{2}}\ket{0}\ket{\phi_0}+\sqrt{\frac{1- D}{2}}e^{i\alpha}\ket{1}\ket{\phi_1}.
\end{eqnarray}
$\ket{0}$ and $\ket{1}$ are single-photon \textit{path states}, i.e. states of well defined trajectory. The parameter $D$ corresponds to the path distinguishability, an attribute associated to the \textit{particle} behaviour of the quanton,  $0\leq D \leq 1$.

For $D=1$, the states $\ket{\psi}$ represent a particle travelling along the path 0.  When $D=0$, there is no distinction between the paths taken by the quanton.  On the other hand, $\ket{\phi_0}$ and $\ket{\phi_1}$ are the polarization states associated to each arm of the interferometer. State $\ket{\phi_1}$ will be expressed in terms of  $\ket{\phi_0}$ and its orthogonal state $\ket{\phi_0^{\perp}}$, which define a basis for the 2-dimensional Hilbert space, as follows
\begin{eqnarray}
\ket{\phi_1}&=&\frac{V}{\sqrt{1-D^2}}\ket{\phi_0}+\frac{C}{\sqrt{1-D^2}}e^{i\beta}\ket{\phi_0^{\perp}}.
\end{eqnarray}
The parameter $V$ correspond to the fringe visibility, associated to wave behaviour of the quanton, and $C$ is the concurrence, a measure of quantum entanglement between both degrees of freedom (path and polarization) of the quanton. $C=0$ corresponds to a product state, while $C=1$ generates maximally entangled states. . 

Notice that, in order to fully specify the state $\ket{\psi}$, five parameters are needed; $D,V$ and $C$, in addition to the phases 
$\alpha$ and $\beta$. The first three parameters are not independent and are related by the equation (\ref{Qian}).

Now, let us specify  the state $|\bar{\psi}\rangle$ of a second quanton, with phases $\bar{\alpha}$ and $\bar{\beta}$, and visibility $\bar{V}$, distinguishability $\bar{D}$ and concurrence $\bar{C}$, as 
\begin{eqnarray}
|\bar{\psi}\rangle&=&\sqrt{\frac{1+\bar{D}}{2}}\ket{0}\ket{\bar{\phi}_0}+\sqrt{\frac{1-\bar{D}}{2}}e^{i\bar{\alpha}}\ket{1}\ket{\bar{\phi}_1},\\
\ket{\bar{\phi}_1}&=&\frac{\bar{V}}{\sqrt{1-\bar{D}^2}}\ket{\bar{\phi}_0}+\frac{\bar{C}}{\sqrt{1-\bar{D}^2}}e^{i\bar{\beta}}\ket{\bar{\phi}_0^{\perp}}.
\end{eqnarray}

Notice that in order to relate both states, two additional parameters $\gamma$ and $\xi$ are needed, defined by 
\begin{eqnarray}
\langle \bar{\phi}_0|\phi_0\rangle=\gamma \text{   and   }
\langle \bar{\phi}^{\perp}_0|\phi^{\perp}_0\rangle=\gamma e^{-i\xi}.
\end{eqnarray}
Without lost of generality the parameter $\gamma$ can be taken to be real and positive, $0\leq \gamma \leq1$.

\section{Bures distance between particle state and an arbitrary state}\label{sec:sec3} 

The magnitude of the inner product between the states $\ket{\psi}$ and $\ket{\bar{\psi}}$ (a measure of the fidelity) is given by 
\begin{eqnarray}\nonumber
|\langle{\bar{\psi}}|\psi\rangle |&=&\Bigg|\frac{\gamma}{2} \frac{\Big[  (1+ D)(1+ \bar{D})+e^{i\lambda_1}V\bar{V}-e^{i\lambda_2}C\bar{C} \Big] }  {\sqrt{(1+ D)(1 +  \bar{D})}} \\ \label{innerproduct}
&+&\frac{\sqrt{1-\gamma^2}}{2} \frac{ \Big[ e^{i\lambda_3}C\bar{V}+e^{i\lambda_4}V\bar{C} \Big] }  {\sqrt{(1+ D)(1 +  \bar{D})}}\Bigg|, 
\end{eqnarray}
where $\lambda_1=\alpha-\bar{\alpha}$, $\lambda_2=\lambda_1+\beta-\bar{\beta}-\xi$, $\lambda_3=\lambda_1+\beta-\xi$, and $\lambda_4=\lambda_1-\bar{\beta}$ (see Appendix \ref{app:A} for the derivation). On the other hand, the \textit{Bures} distance between these states is given by
\begin{eqnarray}\label{BuresDistance}
d_B(\ket{\bar{\psi}},\ket{\psi})=\sqrt{2}\sqrt{1-|\langle{\bar{\psi}}|\psi\rangle |}.
\end{eqnarray}
If we fix the state of the first quanton and allow the second quanton to be any arbitrary state, it is clear that in general the distance between the first (fixed) quanton and the second (arbitrary) quanton is a function of the four parameters  that define the second state ($\bar{V},\bar{D},\bar{\alpha}$ and $\bar{\beta}$) and the two other additional parameters $\gamma$ and $\xi$ needed to connect the first and the second quantons. So, 
\begin{eqnarray}
d_B(\ket{\bar{\psi}},\ket{\psi})=f(\bar{D},\bar{V},\bar{\alpha},\bar{\beta},\gamma, \xi)
\end{eqnarray}

\begin{figure}
 \centering \includegraphics[width=\linewidth]{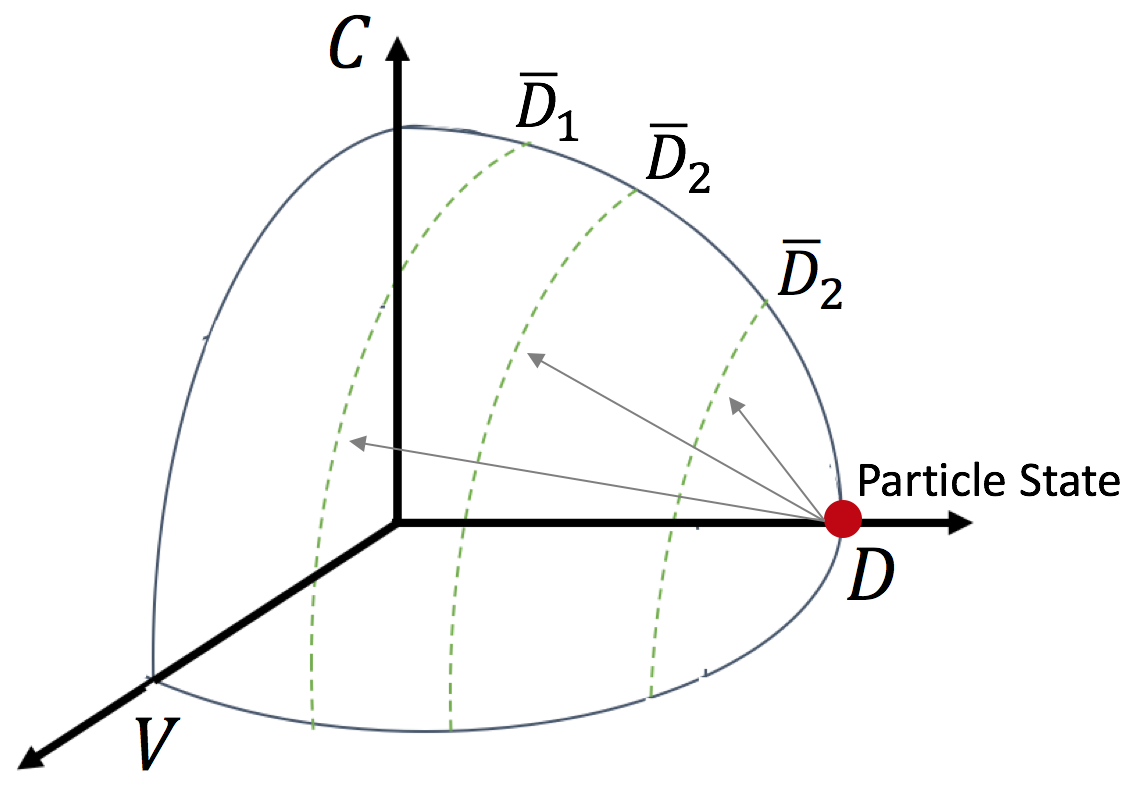}
 \caption{Distance between a general particle state (red dot) and an arbitrary quanton with distinguishability $\bar{D}$. The distance increases as the quantum moves towards the plane of zero which-way information. } \label{fig:fig1}
\end{figure}

However, it is clear from expression (\ref{innerproduct}) that only when the first quanton is a particle state ($D=1$), the distance becomes a simple function of two parameters, $\gamma$ and $\bar{D}$, namely
\begin{eqnarray}\label{distance}
d_B(\ket{\bar{\psi}},\ket{\psi})=\sqrt{2}\sqrt{1-\gamma\sqrt{\frac{1+\bar{D}}{2}}}
\end{eqnarray}
Expression (\ref{distance}) shows that the distance between a  particle state and an arbitrary state with distinguishability $\bar{D}$, visibility $\bar{V}$ and concurrence $\bar{C}$ is independent of the last two parameters.  This special feature is not shared by any other state and is represented in figure (\ref{fig:fig1}), using the VDC sphere.

\begin{figure}
 \centering \includegraphics[width=\linewidth]{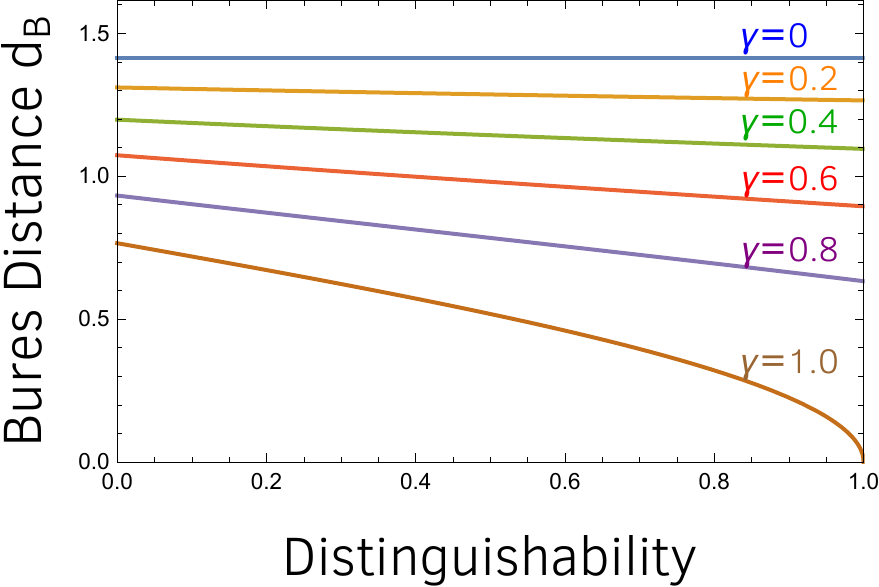}
 \caption{Bures distance between a particle state and a general quanton of distinguishability $D$, for different values of $\gamma$.} \label{fig:fig2}
\end{figure}

Notice that, as the distinguishability increases, the distance becomes smaller, achieving its minimum value of $\sqrt{2}\sqrt{1-\gamma}$ when the second state has perfect distinguishability. On the other hand, when $\bar{D}=0$, i.e. any state exhibiting wave and entanglement properties (but not a well defined trajectory), the distance is maximum and equal to $\sqrt{2}\sqrt{1-\gamma/\sqrt{2}}$. In particular, when the  polarization in the arm 0 of the interferometer is the same for both quantons ($\gamma=1$), then the distances varies from $\sqrt{2-\sqrt{2}}$ to $0$ as $\bar{D}$ increases. On the other hand, if both polarizations are orthogonal, then the distance is constant. This behaviour is represented in figure (\ref{fig:fig2}).

\section{Discussion}\label{sec:sec4} 

From (\ref{BuresDistance}), it is clear that the  distance from a quanton, in a general state $\ket{\bar{\psi}}$, to a particle state $\ket{\psi}$ is minimized for $\gamma=1$, i.e. when the polarization state of the  particle and the polarization state associated to the arm 0 of the quanton, are the same. If we define $\mathcal{P}$ as the set of all particle states in $\mathcal{H}$, then it is clear that

\begin{eqnarray}
\min_{\ket{\psi}\in\mathcal{P}} d_{Bures}(\ket{\psi},\ket{\bar{\psi}})=\sqrt{1-\sqrt{\frac{1+\bar{D}}{2}}}
\end{eqnarray}

\begin{figure}
 \centering \includegraphics[width=\linewidth]{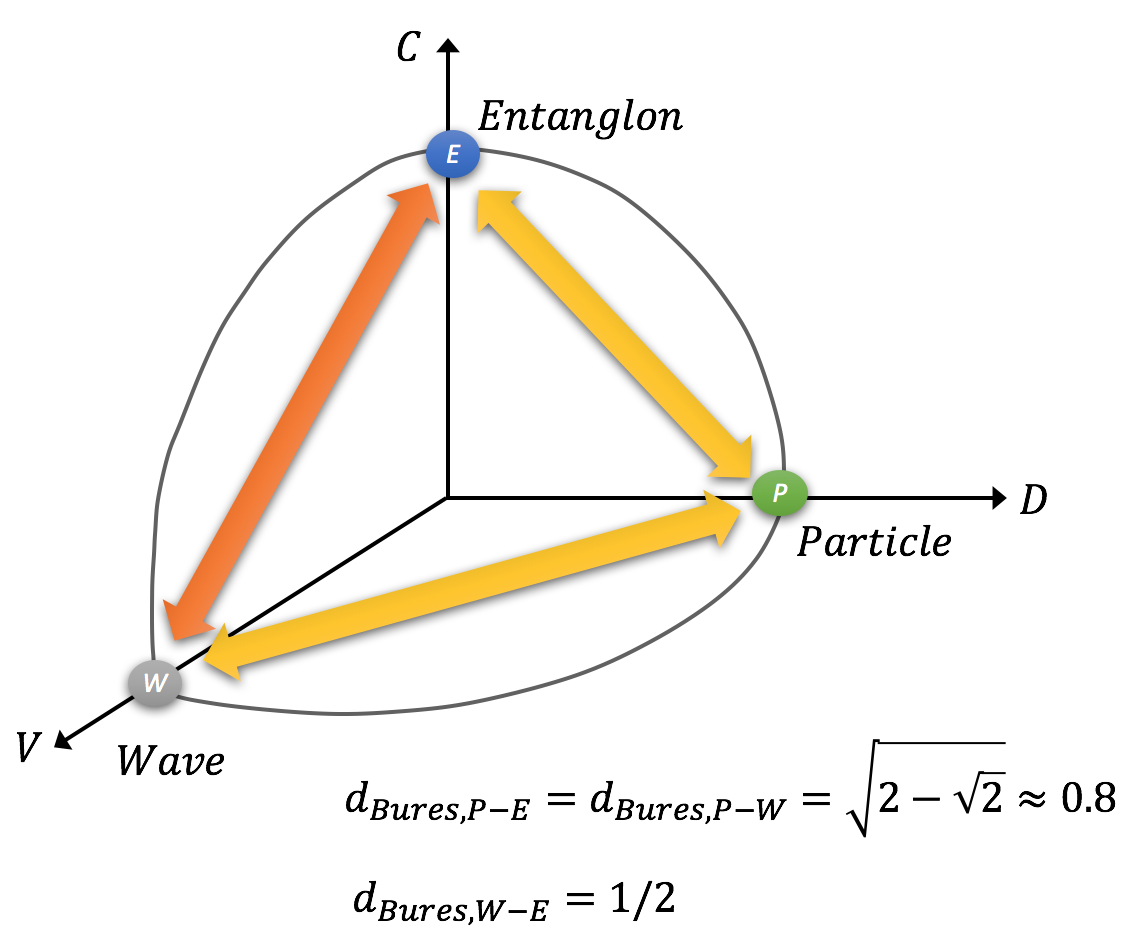}
 \caption{The Bures distance between a Particle ($P$) and a Wave ($W$) is equal to the distance between the Particle and the Entanglon ($E$). However, the distance between the Entanglon and the Wave is not the same as the distance between the Entanglon and the Particle. This example illustrates the special role of the particle states.  } \label{fig:fig3}
\end{figure}

This result shows that \textit{particle states play a central role, different from other states, such as the wave or the entanglon}. For example, the \textit{Bures} distance between an entanglon, $C=1$, and an arbitrary state $\ket{\bar{\psi}}$ depends on its $\bar{C}$ and $\bar{D}$ values, and several relative phases, as can bee seen from (\ref{innerproduct}). The same happens for a wave.  This is can be seen in figure (\ref{fig:fig3}), which shows that $d_{Bures,Particle-Wave}=d_{Particle_Entanglon}\approx0.8$ but $d_{Bures,Entanglon-Wave}\neq d_{Bures,Entanglon-Particle}$.

Therefore, the distinguishability of a general quanton is related to the \textit{Bures} distance between its quantum state and the \textit{nearest particle state}, through an injective function. Hence, the minimum \textit{Bures} distance from a particle state to a generic quantum state quantifies the distinguishability of latter.

%

\section*{ACKNOWLEDGMENTS}
We thank the financial support of Conicyt with the project Fondecyt $\#$1180175.

\appendix

\section{Appendix 1}\label{app:A}
Consider the inner product 
\begin{eqnarray}\nonumber
\langle \bar{\psi} | \psi\rangle = \sqrt{\frac{(1+ \bar{D})(1+ D)}{2}}\langle \bar{\phi}_0 | \phi_0 \rangle\\ \label{Eq1}
+e^{i\lambda_1}\sqrt{\frac{(1- \bar{D})(1- D)}{2}}\langle \bar{\phi}_1 | \phi_1\rangle 
\end{eqnarray}
where $\lambda_1=\alpha-\bar{\alpha}$.  The other products are given by the expressions
\begin{eqnarray}\label{Eq2}
\langle \bar{\phi}_0 | \phi_0 \rangle=\gamma, 
\end{eqnarray}
and
\begin{eqnarray}\nonumber
\langle \bar{\phi}_1 | \phi_1 \rangle=\gamma 
\frac{ \Big(
 V\bar{V}-C\bar{C}e^{i (\beta-\bar{\beta}-\xi)  } \Big)}{\sqrt{(1-D^2)(1-\bar{D}^2)}}+\\ \label{Eq3}
\sqrt{1-\gamma^2} \frac{ \Big(\bar{V}Ce^{i(\beta-\xi)}+V\bar{C}e^{i\bar{\beta}}\Big) }{
 \sqrt{(1-D^2)(1-\bar{D}^2)}}.
\end{eqnarray}
Inserting (\ref{Eq2}) and (\ref{Eq3}) into (\ref{Eq1}), equation (\ref{innerproduct}) is recovered.

\end{document}